\documentclass[onecolumn,showpacs,preprintnumbers,amsmath,amssymb]{revtex4}

\usepackage{graphicx}
\usepackage{color}
\usepackage[latin1]{inputenc}
\usepackage{dcolumn}
\usepackage{bm}
\usepackage{epsfig}

\begin{document}

\preprint{\bf Draft version 9}

\title{Sequential motif profile of natural visibility graphs}

\author{Jacopo Iacovacci, Lucas Lacasa}
\email{j.iacovacci@qmul.ac.uk; l.lacasa@qmul.ac.uk}
\affiliation{School of Mathematical Sciences, Queen Mary University of London, Mile End Road, E14NS London (UK)}%


\begin{abstract}
The concept of sequential visibility graph motifs -subgraphs appearing with characteristic frequencies in the visibility graphs associated to time series- has been advanced recently along with a theoretical framework to compute analytically the motif profiles associated to Horizontal Visibility Graphs (HVGs). Here we develop a theory to compute the profile of sequential visibility graph motifs in the context of Natural Visibility Graphs (VGs). This theory gives exact results for deterministic aperiodic processes with a smooth invariant density or stochastic processes that fulfil the Markov property and have a continuous marginal distribution. The framework also allows for a linear time numerical estimation in the case of empirical time series. A comparison between the HVG and the VG case (including evaluation of their robustness for short series polluted with measurement noise) is also presented.
\end{abstract}


\maketitle

\section{Introduction}

In recent years different methods \cite{Small, Thurner2007, Small2, donner2010, donner2011} have been proposed to map the structure and underlying dynamics of a given time series into an associated graph representation, with the aims of exploiting the modern tools of network science \cite{bollobas2013modern,boccaletti2006complex,newman2003structure} in the traditional task of time series analysis  \cite{pollock1999handbook,box2015time}, thereby building a bridge between the two fields. \\
\noindent In this context, visibility graphs have been proposed \cite{lacasa2008time,luque2009horizontal} as a tool to extract a graph from the relative positions of an ordered series, from which several graph features can be extracted and used for description and classification problems. Very recently we have advanced the concept of sequential visibility graph motifs \cite{iacovacci2015visibility}, building on the idea of network motifs \cite{milo2002network,alon2007network} to explore the decomposition of visibility graphs into sequentially restricted subgraphs. These motifs induce a graph-theoretical symbolization of a given time series into a sequence of subgraphs. We have shown that the marginal distribution of the motif sequence -the so-called motif profile- is an informative feature to describe different types of dynamics and is useful in the task of classifying empirical time series. For large classes of dynamical systems, we were able to develop a theory to analytically compute the frequency of each motif when these are extracted from a so-called Horizontal Visibility Graph (HVG), this being a modified and simpler version of the original (natural) visibility graph (VG) which has often shown analytical tractability \cite{lacasa2014on,pre2013,jpa2014,jns}. As a matter of fact, in the case of VGs to obtain analytical insight has shown to be a challenging task, and besides few exception \cite{epl} most of the works that make use of this statistic are computational. In this paper we bridge this gap and advance a theory to analytically compute the complete motif profile in the natural case (VG motifs). We focus on motifs of size $n=4$ as this was shown to be the simplest case which gives nontrivial results \cite{iacovacci2015visibility}. We validate this theory by deriving explicit motif profiles for several classes of dynamics which we show to be in good agreement with numerical simulations. We also study the robustness of this feature when the time series is short and polluted with measurement noise, and compare its performance with the case of HVG motifs. \\

\noindent The rest of the paper is as follows:    
after recalling the definitions of natural and horizontal visibility graphs, in Section II we present the concept and main properties of sequential visibility graph motifs, as well as recalling the theoretical framework where the motif profile from the horizontal version was derived. In Section III we focus on natural visibility and develop the theory to compute analytically the motif profile associated to processes where the dynamics are either bounded or unbounded. We test this theory by assessing the predictions for different dynamical systems, and we also show that white noise with different marginals can be distinguished using the natural version instead of the horizontal one. In Section IV we show that the visibility graph motif profile is a robust feature in the sense of (i) having a fast convergence to asymptotic values for short series size $N$ and (ii) being robust against contamination with measurement noise (white and colored). 
In Section V we conclude.

\section{Visibility graphs and sequential motifs} 


\begin{figure}
\includegraphics[width= 10 cm]{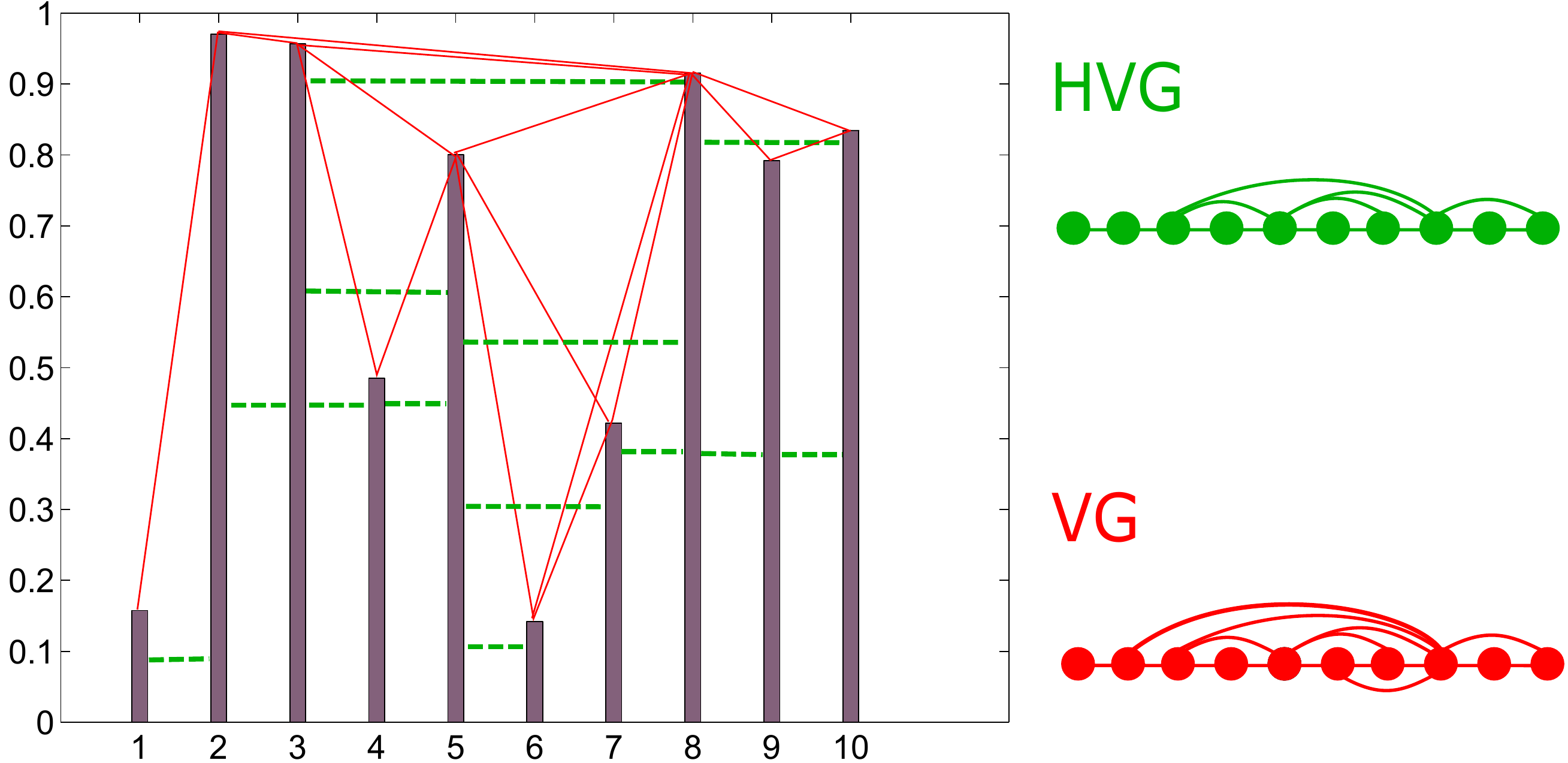}
\caption{(Color online) Two visibility algorithms, Natural Visibility (red) and Horizontal Visibility (green) applied to a time series of 10 data (bars); the corresponding Visibility Graph (VG) and Horizontal Visibility Graph (HVG) are shown on the right: each datum in the series corresponds to a node in the graph and two nodes are connected if their corresponding data heights show respectively natural visibility or horizontal visibility (see the text).}
\label{fig:0}
\end{figure}

Let ${\bf x}=\{x_1,...,x_N\}$ be a real-valued time series of $N$ data. The natural visibility graph (VG) \cite{lacasa2008time} extracted from the series is the graph ${\cal G}=\{V,E\}$ where each datum $x_i$ in the series is associated to a node $v_i$ (thus $|V|=N$ and $V$ is a totally ordered set) and an edge $e_{i,j}\in E$ between node $i$ and node $j$ exists if $x_k< x_i + \frac{k-i}{j-i}[x_j-x_i]$ for each $k$ such that $i<k<j$. This is called the visibility criterion, which for the natural version is indeed a convexity criterion. Analogously, the horizontal visibility graph (HVG) \cite{luque2009horizontal} extracted from the series is the graph ${\cal G}^h=\{V,E^h\}$, with the same vertex set than $\cal G$ and a smaller edge set $E^h$, where an edge $e_{i,j}\in E$ exists between nodes $i$ and $j$ if $x_k< \max(x_i,x_j), \ \forall k: i<k<j$. This visibility criterion is in turn an ordering one. It was indeed shown that the HVG associated to a time series is invariant under monotonic transformations in the series \cite{RW_visibility}, thus the HVG is an order statistic of the series. ${\cal G}^h$ is indeed a non crossing graph \cite{severini} which by construction is always also a sub-graph of ${\cal G}$ (although ${\cal G}$ is not in general planar). Both VG and HVG are connected graphs with a trivial Hamiltonian path given by the sequence of vertices $(v_1,v_2,\dots,v_N)$. An illustration of how to construct a VG and HVG from a given time series is shown in figure \ref{fig:0}.\\

\begin{figure}[h!]
\centering
\includegraphics[width= 12 cm]{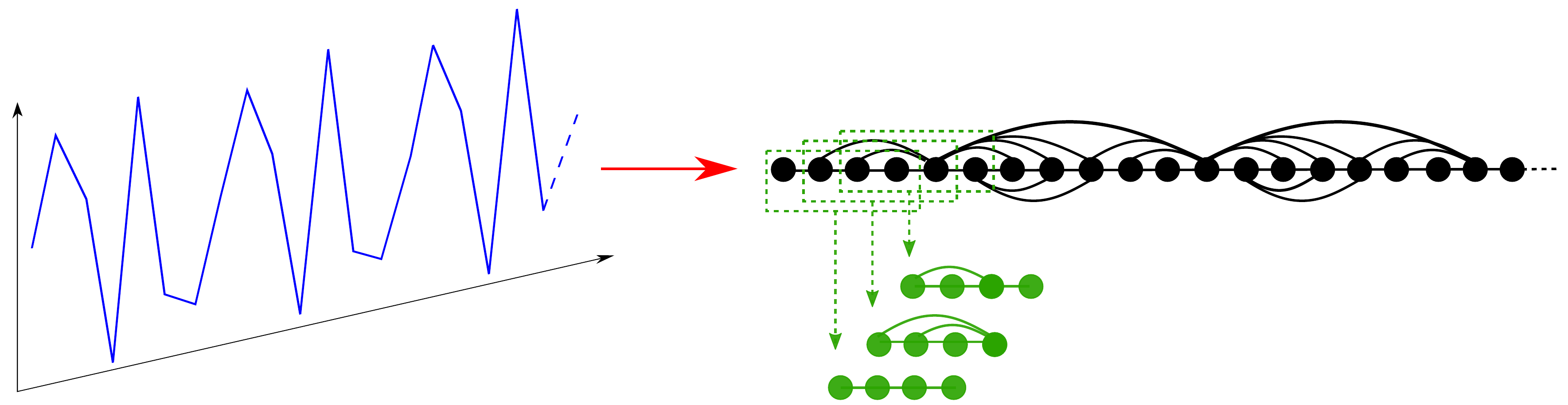}
\caption{Schematic of visibility graph motif detection. A time series is converted into a visibility graph according to the visibility criterion (red arrow). A window of size $n=4$ slides along the Hamiltonian path of the VG graph and detects at each step a different VG motif.}
\label{fig:1}
\end{figure}

\noindent The set of sequential VG motifs of size $n$, $n\in[2,3,...,N]$ is defined as the set of all the $M_{n}$ possible sub-graphs with $n$ consecutive vertices along the Hamiltonian path of a VG (similarly, the set of HVG motifs of size $n$ is the set of all the $M^h_{n}$ admissible sub-graphs with $n$ consecutive vertices along the Hamiltonian path of a HVG). Accordingly, sequential VG motifs are also visibility graphs. For $n=4$, there are in principle a total of $M_{4}=8$ possible motifs (see table \ref{tab:1} for an enumeration), although as we will show below the number of admissible ones is just 6.
Given a VG ${\cal G}$, its sequential motifs can be detected using a sliding window of size $n$ which slides along the Hamiltonian path of the graph with $N-n$ consecutive overlapping steps. At each step a particular motif is detected inside the window. We can accordingly estimate $\Phi_m$, the frequency of appearance of a certain motif $m$, and define the {\em n-motif profile} $Z^n=(\Phi_{1},\dots,\Phi_{M_{n}})$. The process of extracting a VG/HVG and its sequential visibility motif set is illustrated in Figure \ref{fig:1} (the concept is analogous for HVG, although the set of admissible motifs is different in both cases).      
Note that since $Z^n$ can be understood as a discrete probability distribution and is therefore a vector with unit norm (we use the $\ell_1$ norm here) $\sum_{m=1}^{M_{n}} \Phi_m=1$, the number of degrees of freedom of $Z^n$ is $M_{n}-1$ (again, as we will see below, in the case considered here it is even more reduced as the number of admissible motifs will be less than 8).\\    
  
\noindent In a recent work \cite{iacovacci2015visibility} we introduced the concept of sequential HVG motifs and advanced a theory to compute in an exact way $Z^4$ in the case of the HVG. It was shown that the $n=4$ motif statistic was useful to discriminate across different types of dynamics. The case of uncorrelated noise was shown to yield a universal motif profile, independent of the marginal distribution of the i.i.d. process and this enabled the definition of randomness test. We also found for some deterministic dynamics some \emph{forbidden motifs}, which represented a persistent characteristic to test the randomness of a process (note that if a motif of size $n$ doesn't occur, then also all the motifs of size $n'>n$ which incorporate that motif won't occur either).
Since VG and HVG $n$ motif profiles are a temporally constrained feature (they are evaluated along $n$ consecutive nodes on the Hamiltonian path) its extraction can be seen as a process of dynamic symbolization. Under this perspective, the relation between HVG motifs of size $n=4$ and the so-called \emph{ordinal patterns} (OPs) \cite{bandt2002permutation} was acknowledged in \cite{iacovacci2015visibility}. $n$-OPs are symbols extracted from a time series representing the possible ranking output of $n$ consecutive data and are extracted from a specific time series by comparing the value of all the set of $n$ consecutive data along the series \cite{bandt2002permutation,bandt2005ordinal,amigo2010permutation}. It was not unexpected to find a link between HVG $n$-motifs and $n$-OPs as HVG is known to be an order statistic, much as OPs. Indeed, in the particular case of a time series for which data don't repeat $P(x_t=x_{t+1})\simeq 0$ there exists a mapping between each appearing HVG motif and a specific set of ordinal patterns \cite{iacovacci2015visibility}; in this scenario the forbidden motifs selected by the horizontal visibility are, in general, set of the so called \emph{forbidden ordinal patterns} \cite{amigo2006order}. Of course both VG and HVG motifs analysis can be applied without requiring any further assumption to time series taking values from finite sets (namely when $P(x_t=x_{t+1})\neq 0$), while the ordinal patterns approach -based uniquely on the ranking statistics- require further assumptions in that case. 
Here we focus in the {\it natural} version of the algorithm and explore VG motifs instead. As VGs are not invariant under monotonic transformations in the series \cite{RW_visibility}, in general they depend on the marginal probability distribution of the time series and are not an order statistic. 
Accordingly, there is no obvious correspondence between $n$-OPs and VG $n$-motifs and both approaches in principle represent two independent symbolization methods that encode temporal information in a different way. 
In what follows we recall the theoretical framework for HVG motifs and in the next section we extend this theory to deal with VG motifs.

\subsection{Theory for the HVG motif profile}

\begin{table}[h!]
\begin{ruledtabular}
\begin{tabular}{ccc}
{\bf Label}&{\bf HVG motif type}&{\bf Inequality set}\\
1&\begin{minipage}{.1\textwidth}
\includegraphics[width= 0.3\textwidth]{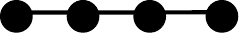}
\end{minipage}&$\{ \forall (x_l,x_{l+1}), x_{l+2}<x_{l+1}, x_{l+3}<x_{l+2}\}\cup\{\forall (x_l,x_{l+3}), x_{l+1}>x_l,x_{l+2}>x_{l+1}\} $\\
2&\begin{minipage}{.1\textwidth}
\includegraphics[width= 0.3\textwidth]{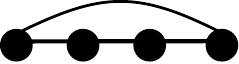}
\end{minipage}& $\{\forall x_l, x_{l+1}<x_l,x_{l+2}=x_{l+1},x_{l+3}>x_{l+2}\}$\\
3&\begin{minipage}{.1\textwidth}
\includegraphics[width= 0.3\textwidth]{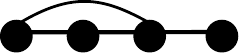}
\end{minipage}& $\{\forall x_l, x_{l+1}<x_l, x_{l+1}<x_{l+2}<x_l,x_{l+3}<x_{l+2}\}\cup\{\forall (x_l,x_{l+3}),x_{l+1}<x_l,x_{l+2}>x_l\}$\\
4&\begin{minipage}{.1\textwidth}
\includegraphics[width= 0.3\textwidth]{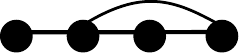}
\end{minipage}& $\{ \forall x_l, x_{l+1}>x_l,x_{l+2}<x_{l+1}, x_{l+3}>x_{l+2}  \}\cup\{\forall x_l, x_{l+1}<x_l, x_{l+2}<x_{l+1}, x_{l+2}<x_{l+3}<x_{l+1}\}$\\
5&\begin{minipage}{.1\textwidth}
\includegraphics[width= 0.3\textwidth]{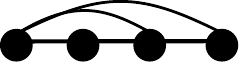}
\end{minipage}&$\{\forall x_l, x_{l+1}<x_l, x_{l+1}<x_{l+2}<x_l, x_{l+3}>x_{l+2}\}$\\
6&\begin{minipage}{.1\textwidth}
\includegraphics[width= 0.3\textwidth]{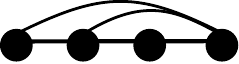}
\end{minipage}& $\{\forall x_l, x_{l+1}<x_l, x_{l+2}<x_{l+1}, x_{l+3}>x_{l+1}\}$\\

\end{tabular}
\end{ruledtabular}
\caption{The set of size-4 HVG motifs are defined according to a set of relations between 4 arbitrary consecutive data $\{x_l,x_{l+1},x_{l+2},x_{l+3}\}$, $l\in[1,N-3]$ in the time series.}
	\label{tab:1_2}
\end{table} 

\noindent Consider a generic dynamical process ${\cal H}:\mathbb{R}\to \mathbb{R}$ with a smooth invariant measure $f(x), \ x\in(a,b)$ that fulfils the Markov property $f(x_l|x_{l-1},x_{l-2},\dots)=f(x_l|x_{l-1})$, where $f(x_l|x_{l-1})$ is the transition probability distribution. It was shown in \cite{iacovacci2015visibility} that, given four arbitrary consecutive data $x_l\dots x_{l+3}$ the motif profile $Z^4=(\Phi^4_1,\dots,\Phi^4_6)$ could then be computed as

\begin{equation}
\Phi^4_{m} = \int f(x_l) dx_l \int f(x_{l+1}|x_l) dx_{l+1} \dots \int f(x_{l+n-1}|x_{l+n-2}) dx_{l+n-1}
\label{eq:HVG}
\end{equation} 
 
where the range of each integral was implicitly given by the inequality set reported in table \ref{tab:1_2}. This inequality set encoded the relative location of data within a particular motif, and indeed highlights the order statistic nature of this measure. In those experimental cases where there is no access to $f(x_t|x_{t-1})$,  
this approach allowed in turn for an numerical estimation of HVG motif profile with linear time complexity $O(N)$.

%
%
%
%
%
%
%


\section{Theory for sequential VG motifs}
Let us consider again a time-discrete (deterministic or stochastic) dynamical process $x_{t+1}={\cal H}(x_t,\xi)$ that fulfils the Markov property: $\forall l \ f(x_l|x_{l-1},x_{l-2},\dots)=f(x_l|x_{l-1})$, where $f(x_l|x_{l-1})$ is the transition probability distribution and $x \in (a,b)$. For deterministic processes $f(x_l|x_{l-1})=\delta(x_l - {\cal H}(x_{l-1}))$ where $\delta(x)$ is the Dirac-delta distribution
where $\delta(x)$ is the Dirac-delta distribution:  
 \begin{equation}
\int_p^q \delta(x-y)dx=
\left\{
\begin{array}{rcl}
     1 & y \in [p,q]\\
     0 & \textrm{otherwise}
\end{array}
\right.
\label{theo}
\end{equation}
and $f(x)$ is a smooth invariant measure of the process ${\cal H}(x)$, whereas for stochastic processes $f(x)$ is simply the underlying probability density, i.e. the marginal distribution of the process. Our theory addresses the motif profile $Z^4$, in what follows we split this analysis in two cases, depend whether $x$ is bounded or unbounded. In both cases, each probability $\Phi^4_m$ is computed formally using concatenated integrals which are formally equivalent to eq.\ref{eq:HVG}, where the ranges of each integral are given according to the convexity criteria defining the visibility rule, as opposed to the HVG case where these were simply ordering criteria. Whereas in the case of unbounded variables the inequality set will only take into account the visibility criteria within the motifs, in the case of bounded variables the additional restriction of variables needing to be bounded adds a layer of complexity as we will see. From now on, let $\{x_l,x_{l+1},x_{l+2},x_{l+3}\}$ be four arbitrary consecutive data ($l\in[1,N-3]$).

\subsection{{\it Unbounded} variable $x\in(-\infty,\infty)$} 

\begin{table}[h!]
\begin{ruledtabular}
\begin{tabular}{ccc}
{\bf Label\quad\quad}&{\bf VG motif type}&{\bf Inequality set}\\
1&\begin{minipage}{.1\textwidth}
\includegraphics[width= 0.3\textwidth]{4motif-1.pdf}
\end{minipage}&$\{ \forall (x_l,x_{l+1}), x_{l+2}<2x_{l+1}-x_l, x_{l+3}<2x_{l+2}-x_{l+1}\}$\\
2&\begin{minipage}{.1\textwidth}
\includegraphics[width= 0.3\textwidth]{4motif-2.pdf}
\end{minipage}& $\{\emptyset\}$ \\
3&\begin{minipage}{.1\textwidth}
\includegraphics[width= 0.3\textwidth]{4motif-3.pdf}
\end{minipage}& $\{\forall (x_l,x_{l+1}), x_{l+2}>2x_{l+1}-x_l,x_{l+3}<\frac{3}{2}x_{l+2}-\frac{1}{2}x_l\}$\\
4&\begin{minipage}{.1\textwidth}
\includegraphics[width= 0.3\textwidth]{4motif-4.pdf}
\end{minipage}& $\{ \forall (x_l,x_{l+1}), x_{l+2}<2x_{l+1}-x_l, 2x_{l+2}-x_{l+1}<x_{l+3}<3x_{l+1}-2x_l\}$\\
5&\begin{minipage}{.1\textwidth}
\includegraphics[width= 0.3\textwidth]{4motif-5.pdf}
\end{minipage}&$\{\forall (x_l,x_{l+1}), x_{l+2}>2x_{l+1}-x_l, \frac{3}{2}x_{l+2}-\frac{1}{2}x_l<x_{l+3}<2x_{l+2}-x_{l+1}\}$\\
6&\begin{minipage}{.1\textwidth}
\includegraphics[width= 0.3\textwidth]{4motif-6.pdf}
\end{minipage}& $\{\forall (x_l,x_{l+1}), x_{l+2}<2x_{l+1}-x_l, x_{l+3}>3x_{l+1}-2x_l\}$\\
7&\begin{minipage}{.1\textwidth}
\includegraphics[width= 0.3\textwidth]{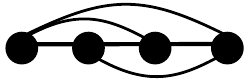}
\end{minipage}& $\{\forall (x_l,x_{l+1}), x_{l+2}>2x_{l+1}-x_l, x_{l+3}>2x_{l+2}-x_{l+1}\}$\\
8&\begin{minipage}{.1\textwidth}
\includegraphics[width= 0.3\textwidth]{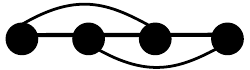}
\end{minipage}& $\{\emptyset\}$\\
\end{tabular}
\end{ruledtabular}
\caption{The set of size-4 VG motifs are defined according to a set of relations between 4 consecutive data $\{x_l,x_{l+1},x_{l+2},x_{l+3}\}$, $l\in[1,N-3]$ in the time  series.}
	\label{tab:1}
\end{table}


In the case of unbounded variables, it is easy to prove that in general
\begin{equation}
\Phi^n_{m} = \int_{\mathbb{R}} f(x_l) dx_l \int^{c^m_1(x_l)}_{d^m_1(x_l)} f(x_{l+1}|x_l) dx_{l+1} \dots \int^{c^m_{n-1}(x_l,\dots,x_{l+n-2})}_{d^m_{n-1}(x_l,\dots,x_{l+n-2})} f(x_{l+n-1}|x_{l+n-2}) dx_{l+n-1}
\label{eq:1}
\end{equation} 
 
where $\{c^m_i(\cdot)\}_{i=1,...,n-1}$ and $\{d^m_i(\cdot)\}_{i=1,...,n-1}$ are the set of functions which specify respectively the upper bound condition and the lower bound condition for the $i$-th integral. As advanced, these conditions are directly related to the visibility criterion (which in the VG case is a convexity relation) and a summary of those are are explicitly reported in table \ref{tab:1} (the equivalence between each motif and its associated inequality set can be proved rigorously although we omit here this proof as it is quite trivial). First, note that for the motifs 2 and 8 the inequality set is empty. This means that these motifs are actually not admissible under the VG algorithm. In the case of motif number 2, note that this motif was an admissible one for HVGs associated to integer-valued series, where the probability of finding equal consecutive data is finite. For VG, it is easy to prove that if the bounding nodes share an edge, then either the left edge or the right edge shares an edge with an inner node. Thus motif 2 is not a VG. Similarly, it is easy to prove that if a time series gives rise to a motif of type 8, then an edge would necessarily appear between the two bounding nodes, reducing this to type 7. Accordingly, the number of admissible motifs is not 8 but 6, and thus the effective number of degrees of freedom associated to $n=4$ VG motifs is $M_n-2-1=5$.\\
To better understand the application of the inequality set in the case of unbounded variables, consider a white Gaussian process ($x_i \in (-\infty,\infty)$) with 
\begin{equation}
f(x_i)=\frac{\exp(-x_i^2/2)}{\sqrt{2\pi}} \  \text{and} \ \quad f(x_{i+1}|x_i)=f(x_{i+1})\nonumber
\end{equation}
the probability of appearance of motif 1 in Table \ref{tab:1} can be written explicitly as
\begin{eqnarray}
\Phi^4_1&=& \int_{-\infty}^{\infty} \frac{e^{-\frac{x_0^2}{2}}}{\sqrt{2\pi}} dx_0 \int_{-\infty}^{\infty} \frac{e^{-\frac{x_1^2}{2}}}{\sqrt{2\pi}} dx_1 \int_{-\infty}^{2x_1-x_0} \frac{e^{-\frac{x_2^2}{2}}}{\sqrt{2\pi}} dx_2 \int_{-\infty}^{2x_2-x1} \frac{e^{-\frac{x_3^2}{2}}}{\sqrt{2\pi}} dx_3 . 
\label{eq:2}
\end{eqnarray}
and the integral can be easily evaluated up to arbitrary precision to obtain $\Phi^4_1\simeq0.13386$.

\subsection{{\it Bounded} variables $x\in[a,b]; \ a,b<\infty$}

In the case where $x\in(a,b)$ where the bounds $a,b <\infty$ are finite, these restrictions in turn induce further conditions on the lower and upper bounds of the integrals in Eq.\ref{eq:1} which have the effect of splitting the overall integral in a sum of different integrals. For illustrative purposes we start by considering a particular example.
Consider a series of i.i.d. uniform random variables $x_i\sim {\cal U}[a,b]$ with $f(x)=(b-a)^{-1}$ and $f(x_{i+1}|x_i)=f(x_{i+1})$, and let us consider again $\Phi_1^4$. According to table \ref{tab:1}, in principle the conditions for the first two variables $x_0$, $x_1$ are $\forall (x_0,x_1)\in[a,b]$; for the third variable $x_2$ the lower bound condition becomes $x_2>a$ but the upper bound condition will depend on the function $2x_1-x_0$ which can take values in $[2a-b,2b-a]$ and thus we need to consider three different cases:     

\begin{equation}
\begin{cases}
     \quad 2x_1-x_0>b  \implies x_1\in (\frac{(x_0+b)}{2},b],\quad x_2\in [a,b]\\
     \quad a<2x_1-x_0<b \implies x_1\in (\frac{(x_0+a)}{2},\frac{(x_0+b)}{2}],\quad x_2\in [a,2x_1-x_0]\\
     \quad 2x_1-x_0<a \quad\quad\quad \{\emptyset\}\\
\end{cases}
\end{equation}
where the last case doesn't contribute (as $x_2>a$ is always fulfilled). Similarly for each admissible choice of the variable $x_2$ the bound conditions for $x_3$ will produce an additional split:   
\begin{equation}
\begin{cases}
     \quad 2x_2-x_1>b  \implies x_2\in (\frac{(x_1+b)}{2},b],\quad x_3\in [a,b]\\
     \quad a<2x_2-x_1<b \implies x_2\in (\frac{(x_1+a)}{2},\frac{(x_1+b)}{2}],\quad x_3\in [a,2x_2-x_1]\\
     \quad 2x_2-x_1<a \quad\quad\quad \{\emptyset\}\\
\end{cases}
\end{equation}
 
After a bit of algebra one finds
\begin{eqnarray}
\Phi^4_1=\frac{1}{(b-a)^{4}} \bigg[\int_{a}^{b} dx_0 \int_{\frac{(x_0+b)}{2}}^{b} dx_1 \int_{\frac{(x_1+b)}{2}}^{b}dx_2 \int_{a}^{b}dx_3+\\ 
+\int_{a}^{b} dx_0 \int_{\frac{(x_0+b)}{2}}^{b} dx_1 \int_{\frac{(x_1+a)}{2}}^{\frac{(x_1+b)}{2}}dx_2 \int_{a}^{2x_2-x_1}dx_3+\\
+\int_{a}^{b} dx_0 \int_{\frac{(2x_0+b)}{3}}^{\frac{(x_0+b)}{2}} dx_1 \int_{\frac{(x_1+b)}{2}}^{2x_1-x_0}dx_2 \int_{a}^{b}dx_3+\\
+\int_{a}^{b} dx_0 \int_{\frac{(2x_0+a)}{3}}^{\frac{(2x_0+b)}{3}} dx_1 \int_{\frac{(2x_1+a)}{2}}^{2x_1-x_0}dx_2 \int_{a}^{2x_2-x_1}dx_3+\\
+\int_{a}^{b} dx_0 \int_{\frac{(2x_0+b)}{3}}^{\frac{(x_0+b)}{2}} dx_1 \int_{\frac{(x_1+a)}{2}}^{\frac{(x_1+b)}{2}}dx_2 \int_{a}^{2x_2-x_1}dx_3\bigg]=\frac{5}{36}\simeq 0.1389.
\label{eq:3}
\end{eqnarray}
Two comments are in order. First, note that this result is different from the value for $\Phi_1^4$ for Gaussian white noise, that is, the results for white noise seems to be dependent on the marginal distribution of the noise. This lack of invariance was expected as VG is not an order statistic, and differs from the phenomenology found for HVG, where results for white noise were universal (independent from the marginal distribution). This evidence will be confirmed in the next sections.
Second, for uniform white noise this result appears independent from the values $a$ and $b$, suggesting that results could depend on the marginal but be independent upon rescaling of the support $(a,b)$. This is however not the case in general, and this independence seems to be unique for uniformly distributed white noise (results not shown).\\
After a bit of algebra, we are able to translate the visibility and bounded variable restrictions inside each motif into another set of inequalities, which we have reported in
table \ref{tab:2}. In what follows we make use of this theory to compute the theoretical VG motif profile for several dynamical processes.

\begin{table}[h!]
\begin{ruledtabular}
\begin{tabular}{ccc}
{\bf Motif label}&{\bf Motif type}&{\bf Inequality set}\\
1&\begin{minipage}{.1\textwidth}
\includegraphics[width= 0.3\textwidth]{4motif-1.pdf}
\end{minipage}&$\{ x_l\in[a,b],x_{l+1}\in[\frac{(x_l+b)}{2},b],x_{l+2}\in[\frac{(x_{l+1}+b)}{2},b], x_{l+3}\in[a,b]\}$\\ 
& & $\{ x_l\in[a,b],x_{l+1}\in[\frac{(x_l+b)}{2},b],x_{l+2}\in[\frac{(x_{l+1}+a)}{2},\frac{(x_{l+1}+b)}{2}], x_{l+3}\in[a,2x_{l+2}-x_{l+1}]\}$\\
& & $\{ x_l\in[a,b],x_{l+1}\in[\frac{(2x_l+b)}{3},\frac{(x_l+b)}{2}],x_{l+2}\in[\frac{(x_{l+1}+b)}{2},2x_{l+1}-x_{l}], x_{l+3}\in[a,b]\}$\\
& & $\{ x_l\in[a,b],x_{l+1}\in[\frac{(2x_l+a)}{3},\frac{(2x_l+b)}{3}],x_{l+2}\in[\frac{(x_{l+1}+a)}{2},2x_{l+1}-x_{l}], x_{l+3}\in[a,2x_{l+2}-x_{l+1}]\}$\\
& & $\{ x_l\in[a,b],x_{l+1}\in[\frac{(2x_l+b)}{3},\frac{(x_l+b)}{2}],x_{l+2}\in[\frac{(x_{l+1}+a)}{2},\frac{(x_{l+1}+b)}{2}], x_{l+3}\in[a,2x_{l+2}-x_{l+1}]\}$\\
2&\begin{minipage}{.1\textwidth}
\includegraphics[width= 0.3\textwidth]{4motif-2.pdf}
\end{minipage}& $\{\emptyset\}$ \\
3&\begin{minipage}{.1\textwidth}
\includegraphics[width= 0.3\textwidth]{4motif-3.pdf}
\end{minipage}&$\{ x_l\in[a,b],x_{l+1}\in[a,\frac{(x_l+b)}{2}],x_{l+2}\in[\frac{(x_{l}+2b)}{3},b], x_{l+3}\in[a,b]\}$\\ 
& & $\{ x_l\in[a,b],x_{l+1}\in[a,\frac{(x_l+a)}{2}],x_{l+2}\in[\frac{(x_{l}+2a)}{3},\frac{(x_{l}+2b)}{3}], x_{l+3}\in[a,\frac{3x_{l+2}-x_{l}}{2}]\}$\\
& & $\{ x_l\in[a,b],x_{l+1}\in[\frac{(2x_l+b)}{3},\frac{(x_l+b)}{2}],x_{l+2}\in[2x_{l+1}-x_{l},b], x_{l+3}\in[a,b]\}$\\
& & $\{ x_l\in[a,b],x_{l+1}\in[\frac{(x_l+a)}{2},\frac{(2x_l+b)}{3}],x_{l+2}\in[\frac{(x_{l}+2b)}{3},b], x_{l+3}\in[a,b]\}$\\
& & $\{ x_l\in[a,b],x_{l+1}\in[\frac{(2x_l+a)}{3},\frac{(2x_l+b)}{3}],x_{l+2}\in[2x_{l+1}-x_{l},\frac{(x_{l}+2b)}{3}], x_{l+3}\in[a,3x_{l+1}-2x_{l}]\}$\\
& & $\{ x_l\in[a,b],x_{l+1}\in[\frac{(x_l+a)}{2},\frac{(2x_l+a)}{3}],x_{l+2}\in[\frac{(x_{l}+2a)}{3},\frac{(x_{l}+2b)}{3}], x_{l+3}\in[a,3x_{l+1}-2x_{l}]\}$\\
4&\begin{minipage}{.1\textwidth}
\includegraphics[width= 0.3\textwidth]{4motif-4.pdf}
\end{minipage}&$\{ x_l\in[a,b],x_{l+1}\in[\frac{(x_l+b)}{2},b],x_{l+2}\in[\frac{(x_{l+1}+a)}{2},\frac{(x_{l+1}+b)}{2}], x_{l+3}\in[2x_{l+2}-x_{l+1},b]\}$\\ 
& & $\{ x_l\in[a,b],x_{l+1}\in[\frac{(x_l+b)}{2},b],x_{l+2}\in[a,\frac{(x_{l+1}+a)}{2}], x_{l+3}\in[a,b]\}$\\
& & $\{ x_l\in[a,b],x_{l+1}\in[\frac{(2x_l+a)}{3},\frac{(2x_l+b)}{3}],x_{l+2}\in[\frac{(x_{l+1}+a)}{2},2x_{l+1}-x_{l}], x_{l+3}\in[2x_{l+2}-x_{l+1},3x_{l+1}-2x_{l}]\}$\\
& & $\{ x_l\in[a,b],x_{l+1}\in[\frac{(2x_l+b)}{3},b],x_{l+2}\in[\frac{(x_{l+1}+a)}{2},\frac{(x_{l+1}+b)}{2}], x_{l+3}\in[2x_{l+2}-x_{l+1},b]\}$\\
& & $\{ x_l\in[a,b],x_{l+1}\in[\frac{(2x_l+a)}{3},\frac{(2x_l+b)}{3}],x_{l+2}\in[a,\frac{(x_{l+1}+a)}{2}], x_{l+3}\in[a,3x_{l+1}-2x_{l}]\}$\\
& & $\{ x_l\in[a,b],x_{l+1}\in[\frac{(2x_l+b)}{3},\frac{(x_l+b)}{2}],x_{l+2}\in[a,\frac{(x_{l+1}+a)}{2}], x_{l+3}\in[a,b]\}$\\
5&\begin{minipage}{.1\textwidth}
\includegraphics[width= 0.3\textwidth]{4motif-5.pdf}
\end{minipage}&$\{ x_l\in[a,b],x_{l+1}\in[a,\frac{(x_l+a)}{2}],x_{l+2}\in[\frac{(x_{l+1}+2a)}{3},\frac{(x_{l+1}+b)}{2}], x_{l+3}\in[\frac{3x_{l+2}-x_{l}}{2},2x_{l+2}-x_{l+1}]\}$\\ 
& & $\{ x_l\in[a,b],x_{l+1}\in[\frac{(2x_l+a)}{3},\frac{(2x_l+b)}{3}],x_{l+2}\in[2x_{l+1}-x_{l},\frac{(x_{l+1}+b)}{2}], x_{l+3}\in[\frac{3x_{l+2}-x_{l}}{2},2x_{l+2}-x_{l+1}]\}$\\
& & $\{ x_l\in[a,b],x_{l+1}\in[\frac{(x_l+a)}{2},\frac{(2x_l+a)}{3}],x_{l+2}\in[\frac{(2x_{l}+a)}{3},\frac{(x_{l+1}+b)}{2}], x_{l+3}\in[\frac{3x_{l+2}-x_{l}}{2},2x_{l+2}-x_{l+1}]\}$\\
& & $\{ x_l\in[a,b],x_{l+1}\in[a,\frac{(x_l+a)}{2}],x_{l+2}\in[\frac{(x_{l+1}+a)}{2},\frac{(x_{l}+2a)}{3}], x_{l+3}\in[a,2x_{l+2}-x_{l+1}]\}$\\
& & $\{ x_l\in[a,b],x_{l+1}\in[\frac{(x_l+a)}{2},\frac{(2x_l+a)}{3}],x_{l+2}\in[\frac{(x_{l+1}+a)}{2},\frac{(x_{l}+2a)}{3}], x_{l+3}\in[a,2x_{l+2}-x_{l+1}]\}$\\
& & $\{ x_l\in[a,b],x_{l+1}\in[a,\frac{(2x_l+b)}{3}],x_{l+2}\in[\frac{(x_{l+1}+b)}{2},\frac{(x_{l}+2b)}{3}], x_{l+3}\in[\frac{3x_{l+2}-x_{l}}{2},b]\}$\\
6&\begin{minipage}{.1\textwidth}
\includegraphics[width= 0.3\textwidth]{4motif-6.pdf}
\end{minipage}&$\{ x_l\in[a,b],x_{l+1}\in[\frac{(2x_l+a)}{3},\frac{(2x_l+b)}{3}],x_{l+2}\in[a,2x_{l+1}-x_{l}], x_{l+3}\in[3x_{l+1}-2x_{l},b]\}$\\ 
& & $\{ x_l\in[a,b],x_{l+1}\in[\frac{(x_l+a)}{2},\frac{(2x_l+a)}{3}],x_{l+2}\in[a,2x_{l+1}-x_{l}], x_{l+3}\in[a,b]\}$\\
7&\begin{minipage}{.1\textwidth}
\includegraphics[width= 0.3\textwidth]{4motif-7.pdf}
\end{minipage}&$\{ x_l\in[a,b],x_{l+1}\in[a,\frac{(x_l+a)}{2}],x_{l+2}\in[a,\frac{(x_{l+1}+a)}{2}], x_{l+3}\in[a,b]\}$\\ 
& & $\{ x_l\in[a,b],x_{l+1}\in[a,\frac{(x_l+a)}{2}],x_{l+2}\in[\frac{(x_{l+1}+a)}{2},\frac{(x_{l+1}+b)}{2}], x_{l+3}\in[2x_{l+2}-x_{l+1},b]\}$\\
& & $\{ x_l\in[a,b],x_{l+1}\in[\frac{(x_l+a)}{2},\frac{(2x_l+a)}{3}],x_{l+2}\in[2x_{l+1}-x_{l},\frac{(x_{l+1}+a)}{2}], x_{l+3}\in[a,b]\}$\\
& & $\{ x_l\in[a,b],x_{l+1}\in[\frac{(x_l+a)}{2},\frac{(2x_l+a)}{3}],x_{l+2}\in[\frac{(x_{l+1}+a)}{2},\frac{(x_{l+1}+b)}{2}], x_{l+3}\in[2x_{l+2}-x_{l+1},b]\}$\\
& & $\{ x_l\in[a,b],x_{l+1}\in[\frac{(2x_l+a)}{3},\frac{(2x_l+b)}{3}],x_{l+2}\in[2x_{l+1}-x_{l},\frac{(x_{l+1}+b)}{2}], x_{l+3}\in[2x_{l+2}-x_{l+1},\frac{(x_{l+1}+b)}{2}]\}$\\
8&\begin{minipage}{.1\textwidth}
\includegraphics[width= 0.3\textwidth]{4motif-8.pdf}
\end{minipage}& $\{\emptyset\}$\\
\end{tabular}
\end{ruledtabular}
\caption{Sets of inequalities between 4 consecutive data $\{x_l,x_{l+1},x_{l+2},x_{l+3}\}$, $l\in[1,N-3]$ in a time series of length $N$ which define the VG motifs of size 4 in the case of bounded variables $x_i\in[a,b]$.}
	\label{tab:2}
\end{table}

 \subsection{VG motif profiles of different families of dynamical systems}

\subsubsection{Fully chaotic logistic map}
We start by considering the fully chaotic logistic map
${\cal H}(x)=4x(1-x), \ x\in [0,1]$, with invariant density $f(x)=\frac{1}{\pi\sqrt{x(1-x)}}$. As this process is deterministic, it fulfils a trivial Markov property such that $f(x_2|x_1)=\delta(x_2-{\cal H}(x_1))$. The HVG motif profile for this process was computed exactly in \citep{iacovacci2015visibility}, here we compute the VG motif profile. Before proceeding to compute each probability contribution, it is important to highlight a subtle point. Since for this process $x \in [0,1]$ is bounded, in principle one should use the inequality set depicted for bounded variables in table \ref{tab:2}. However, in this particular case it is actually not necessary to explicitly consider the restriction $x \in [0,1]$. As we will see in a moment, this is already taken into account implicitly in the computation of each integral and therefore one can use the (simpler) inequality set for unbounded variables given in table \ref{tab:1}.\\
 
\noindent We start by computing $\Phi^4_1$:
\begin{eqnarray}
\Phi^4_1 = \int_0^1 f(x_0)dx_0 \int_0^1 \delta (x_1 - {\cal H}(x_0)) dx_1 \int_0^{2x_1-x_0} \delta(x_2- {\cal H}^2(x_0) )dx_2\int_{0}^{2x_2-x1} \delta(x_3-{\cal H}^3(x0))dx_3 \nonumber 
\end{eqnarray}
which gives the following conditions:\\
${\cal H}^3(x_0)< 2{\cal H}^2(x_0)-{\cal H}(x_0)$\\
${\cal H}^2(x_0)< 2{\cal H}(x_0)-x_0$\\
which are satisfied for $x_0 \in [0.1743,0.25]$. Note at this point that the latter conditions are also satisfied in other ranges, but we only consider those ranges that belong to $[0,1]$, and this is indeed the reason why we don't need to use in this case the inequality set for bounded variables. We thus have
$$\Phi^4_1 \simeq\frac{1}{\pi} B_{\left[0.1743,0.25\right]}\left(\frac{1}{2},\frac{1}{2}\right)\simeq0.0591$$

\noindent As $\Phi^4_2=0$ by construction, we proceed by calculating $\Phi^4_3$: 
\begin{eqnarray}
\Phi^4_3 = \int_0^1 f(x_0)dx_0 \int_0^1 \delta (x_1 - {\cal H}(x_0)) dx_1 \int_{2x_1-x_0}^{1} \delta(x_2- {\cal H}^2(x_0) )dx_2\int_{0}^{\frac{3}{2}x_2-\frac{1}{2}x_0} \delta(x_3-{\cal H}^3(x_0))dx_3 \nonumber 
\end{eqnarray}
which gives the following conditions:\\
${\cal H}^3(x_0)< \frac{3}{2}{\cal H}^2(x_0)-\frac{1}{2}{\cal H}(x_0)$\\
${\cal H}^2(x_0)> 2{\cal H}(x_0)-x_0$\\
which are satisfied for $x_0 \in [0.0522,0.1743]\cup[0.75,0.929]$. Therefore:
$$\Phi^4_3 \simeq\frac{1}{\pi} (B_{\left[0.0522,0.1743\right]}\left(\frac{1}{2},\frac{1}{2}\right)+B_{\left[0.75,0.929\right]}\left(\frac{1}{2},\frac{1}{2}\right))\simeq0.289.$$
Similarly for $\Phi^4_4$ we have
\begin{eqnarray}
\Phi^4_4 = \int_0^1 f(x_0)dx_0 \int_0^1 \delta (x_1 - {\cal H}(x_0)) dx_1 \int_{0}^{2x_1-x_0} \delta(x_2- {\cal H}^2(x_0) )dx_2\int_{2x_2-x_1}^{3x_1-2x_0} \delta(x_3-{\cal H}^3(x_0))dx_3 \nonumber 
\end{eqnarray}
which gives the following conditions:\\
$2{\cal H}^2(x_0)-{\cal H}(x_0)<{\cal H}^3(x_0)< 3{\cal H}^2(x_0)-2x_0$\\
${\cal H}^2(x_0)< 2{\cal H}(x_0)-x_0$\\
which are satisfied for $x_0 \in [0.25,0.75]$, and therefore 
$$\Phi^4_4 \simeq\frac{1}{\pi} B_{\left[0.25,0.75\right]}\left(\frac{1}{2},\frac{1}{2}\right)\simeq0.3333$$
For $\Phi^4_5$ we have
\begin{eqnarray}
\Phi^4_5 = \int_0^1 f(x_0)dx_0 \int_0^1 \delta (x_1 - {\cal H}(x_0)) dx_1 \int_{2x_1-x_0}^{1} \delta(x_2- {\cal H}^2(x_0) )dx_2\int_{\frac{3}{2}x_2-\frac{1}{2}x_0}^{2x_2-x_1} \delta(x_3-{\cal H}^3(x_0))dx_3 \nonumber 
\end{eqnarray}
which gives the following conditions:\\
$\frac{3}{2}{\cal H}^2(x_0)-\frac{1}{2}x_0<{\cal H}^3(x_0)< 2{\cal H}^2(x_0)-{\cal H}(x_0)$\\
${\cal H}^2(x_0)> 2{\cal H}(x_0)-x_0$\\
which are satisfied for $x_0 \in [0.927,0.954]$, and thus
$$\Phi^4_5 \simeq\frac{1}{\pi} (B_{\left[0.04568,0.05224\right]}\left(\frac{1}{2},\frac{1}{2}\right)+B_{\left[0.9239,0.9543\right]}\left(\frac{1}{2},\frac{1}{2}\right)\simeq0.0439$$
For $\Phi^4_6$ we have
\begin{eqnarray}
\Phi^4_6 = \int_0^1 f(x_0)dx_0 \int_0^1 \delta (x_1 - {\cal H}(x_0)) dx_1 \int_{0}^{2x_1-x_0} \delta(x_2- {\cal H}^2(x_0) )dx_2\int_{3x_1-2x_0}^{1} \delta(x_3-{\cal H}^3(x_0))dx_3 \nonumber 
\end{eqnarray}
which gives the following conditions:\\
${\cal H}^3(x_0)> 3{\cal H}(x_0)-2x_0$\\
${\cal H}^2(x_0)< 2{\cal H}(x_0)-x_0$\\
which are never satisfied and thus
$$\Phi^4_6 =0.$$
For $\Phi^4_7$ we have
\begin{eqnarray}
\Phi^4_7 = \int_0^1 f(x_0)dx_0 \int_0^1 \delta (x_1 - {\cal H}(x_0)) dx_1 \int_{2x_1-x_0}^{1} \delta(x_2- {\cal H}^2(x_0) )dx_2\int_{2x_2-x_1}^{1} \delta(x_3-{\cal H}^3(x_0))dx_3 \nonumber 
\end{eqnarray}
which gives the following conditions:\\
${\cal H}^3(x_0)> 2{\cal H}^2(x_0)-{\cal H}(x_0)$\\
${\cal H}^2(x_0)> 2{\cal H}(x_0)-x_0$\\
which are satisfied for $x_0 \in [0,0.0457]\cup[0.9544,1]$, and thus
$$\Phi^4_7 \simeq\frac{1}{\pi} (B_{\left[0,0.046\right]}\left(\frac{1}{2},\frac{1}{2}\right)+B_{\left[0.95,1\right]}\left(\frac{1}{2},\frac{1}{2}\right))\simeq0.2741$$
Finally, by construction $\Phi^4_8=0$. Altogether, we find the VG motif profile of a fully chaotic logistic map
\begin{eqnarray}
{\bf Z}^4=(0.0591,0,0.289, 0.3333, 0.0439, 0, 0.2741,0)
\label{log}
\end{eqnarray}
Note that while the result is in this case an approximation, our theory allows for numerical estimates with arbitrary precision (the result is not exact because the location of fixed points of the map is only approximate, although this approximation is arbitrarily close to the true values).

\subsubsection{Uniform white noise}
For white uniform noise $x(t)=\xi$, $\xi \sim U[a,b]$ we have a probability density $f(x)$ and transition probability $f(x_2|x_1)$ given by 
\begin{equation}
f(x)=\frac{1}{b-a}, \  \text{and} \ \quad f(x_2|x_1)=f(x_2)
\end{equation}
In this case the computations are more cumbersome since we need to make use of the inequality set for bounded variables described in table \ref{tab:2}. The $m$th component of ${\bf Z}^4$ is given by 
\begin{eqnarray}
\Phi^4_m&=&\sum_s \int_{-\infty}^{\infty} \frac{e^{-\frac{x_0^2}{2}}}{\sqrt{2\pi}} dx_0 \int_{d_{s1}^m}^{c_{s1}^m} \frac{e^{-\frac{x_1^2}{2}}}{\sqrt{2\pi}} dx_1 \int_{d_{s2}^m}^{c_{s2}^m} \frac{e^{-\frac{x_2^2}{2}}}{\sqrt{2\pi}} dx_2 \int_{d_{s3}^m}^{c_{s3}^m} \frac{e^{-\frac{x_3^2}{2}}}{\sqrt{2\pi}} dx_3 \nonumber\\
\label{}
\end{eqnarray}
where the sum runs over all the set $s$ of conditions ${d_{si}^m,c_{si}^m}$ which contribute to evaluate the probability for motif $m$ in Table \ref{tab:2}. All the integrals can nonetheless be solved analytically in closed form and give the following motif profile 
\begin{eqnarray}
{\bf Z}^4=\bigg(\frac{5}{36},0, \frac{31}{108}, \frac{31}{108}, \frac{2}{27}, \frac{2}{27}, \frac{5}{36},0\bigg)
\end{eqnarray}
Several comments are in order. First, this profile is different from the one found for the chaotic logistic map. Second, this motif profile turns to be independent from the bounding values $a$ and $b$ (where $x\in[a,b]$), meaning that white, uniform noise has a VG motif profile which is invariant under various transformations in the original distribution of the time series. This is not a trivial property and is indeed a peculiarity of the uniform distribution, in other words the VG motif profile of white noise extracted a from bounded distribution \textit{generally} depends on the bounds of the distribution. 

\subsubsection{Gaussian white noise}
For standard white Gaussian noise $x(t)=\xi$, $\xi \sim {\cal N}(0,1)$ the probability density $f(x)$ and transition probability $f(x_2|x_1)$ are given by 
\begin{equation}
f(x)=\frac{\exp(-x^2/2)}{\sqrt{2\pi}}, \  \text{and} \ \quad f(x_2|x_1)=f(x_2)
\end{equation}
The $m$ component of ${\bf Z}^4$ is given by 
\begin{eqnarray}
\Phi^4_m&=& \int_{-\infty}^{\infty} \frac{e^{-\frac{x_0^2}{2}}}{\sqrt{2\pi}} dx_0 \int_{-\infty}^{\infty} \frac{e^{-\frac{x_1^2}{2}}}{\sqrt{2\pi}} dx_1 \int_{d_2^m}^{c_2^m} \frac{e^{-\frac{x_2^2}{2}}}{\sqrt{2\pi}} dx_2 \int_{d_3^m}^{c_3^m} \frac{e^{-\frac{x_3^2}{2}}}{\sqrt{2\pi}} dx_3 \nonumber\\
\label{}
\end{eqnarray}
where $(d_i^m,c_i^m)$ are now the top and bottom conditions for the variable $x_{l+i}$ in motif $m$ reported in Table for unbounded variables \ref{tab:1}. The integrals can be evaluated numerically up to arbitrary precision and they give the following results
\begin{eqnarray}
{\bf Z}^4=(0.13386,0,0.2850, 0.2850, 0.0811, 0.0811, 0.13386,0)
\end{eqnarray}
At odds with what happens for HVG motifs \citep{iacovacci2015visibility}, this result is different from the benchmark result for uniformly distributed white noise, thus there is not a universal VG motif profile for white noise as previously anticipated.

\subsubsection{Gaussian red noise}

Gaussian colored (red) noise with exponentially decaying correlations can be simulated using an $AR(1)$ process:
\begin{equation}    
x_{t}= r x_{t-1}+\xi
\label{eq:rednoise}
\end{equation} 
where $\xi \sim {\cal N}(0,1)$ is Gaussian white, and $0<r< 1$ is a parameter that tunes the correlation. The auto-correlation function $C(t)$ decays exponentially $C(t)=e^{-t/\tau}$, where the characteristic time  $\tau=1/\ln(r)$.
This model is Markovian and stationary, with a probability density $f(x)$ and transition probability $f(x_2|x_1)$ given by
\begin{equation}
f(x)=\frac{\exp(-x^2/2)}{\sqrt{2\pi}}, \  \text{and} \ f(x_2|x_1)=\frac{\exp[-(x_2-rx_1)^2/(2(1-r^2))]}{\sqrt{2\pi (1-r^2)}}
\end{equation}

\noindent The $m$ component of ${\bf Z}^4$ is given by 
\begin{eqnarray}
\Phi^4_m&=& \int_{-\infty}^{\infty} \frac{e^{\frac{-x_0^2}{2}}}{\sqrt{2\pi}} dx_0 \int_{-\infty}^{\infty} \frac{e^{\frac{-(x_1-rx_0)^2}{2(1-r^2)}}}{\sqrt{2\pi (1-r^2)}} dx_1 \int_{d_2^m}^{c_2^m} \frac{e^{\frac{-(x_2-rx_1)^2}{2(1-r^2)}}}{\sqrt{2\pi (1-r^2)}} dx_2 \int_{d_3^m}^{c_3^m} \frac{e^{\frac{-(x_3-rx_2)^2}{2(1-r^2)}}}{\sqrt{2\pi (1-r^2)}} dx_3
\label{}
\end{eqnarray}
where, again, $(d_i^m,c_i^m)$ are the top and bottom conditions for the variable $x_{l+i}$ in motif $m$ reported in table Table \ref{tab:1}.
Once set the parameter $r$ the profile can be evaluated numerically up to arbitrary precision; here we give the profile for three possible values $r=1/4,1/2$ and $3/4)$ 
\begin{eqnarray}
{\bf Z}^4_{r=\frac{1}{4}}=(0.14713,0,0.27028, 0.27028, 0.08259, 0.08259,
 0.14713,0)\nonumber\\
{\bf Z}^4_{r=\frac{1}{2}}=(0.15731,0, 0.2595, 0.2595, 0.08316, 0.08316,
 0.15731,0)\\
{\bf Z}^4_{r=\frac{3}{4}}=(0.16410,0, 0.25258, 0.25258, 0.08332, 0.08332,
 0.16410,0)\nonumber
\end{eqnarray} 

\begin{figure}
\includegraphics[width= 9cm]{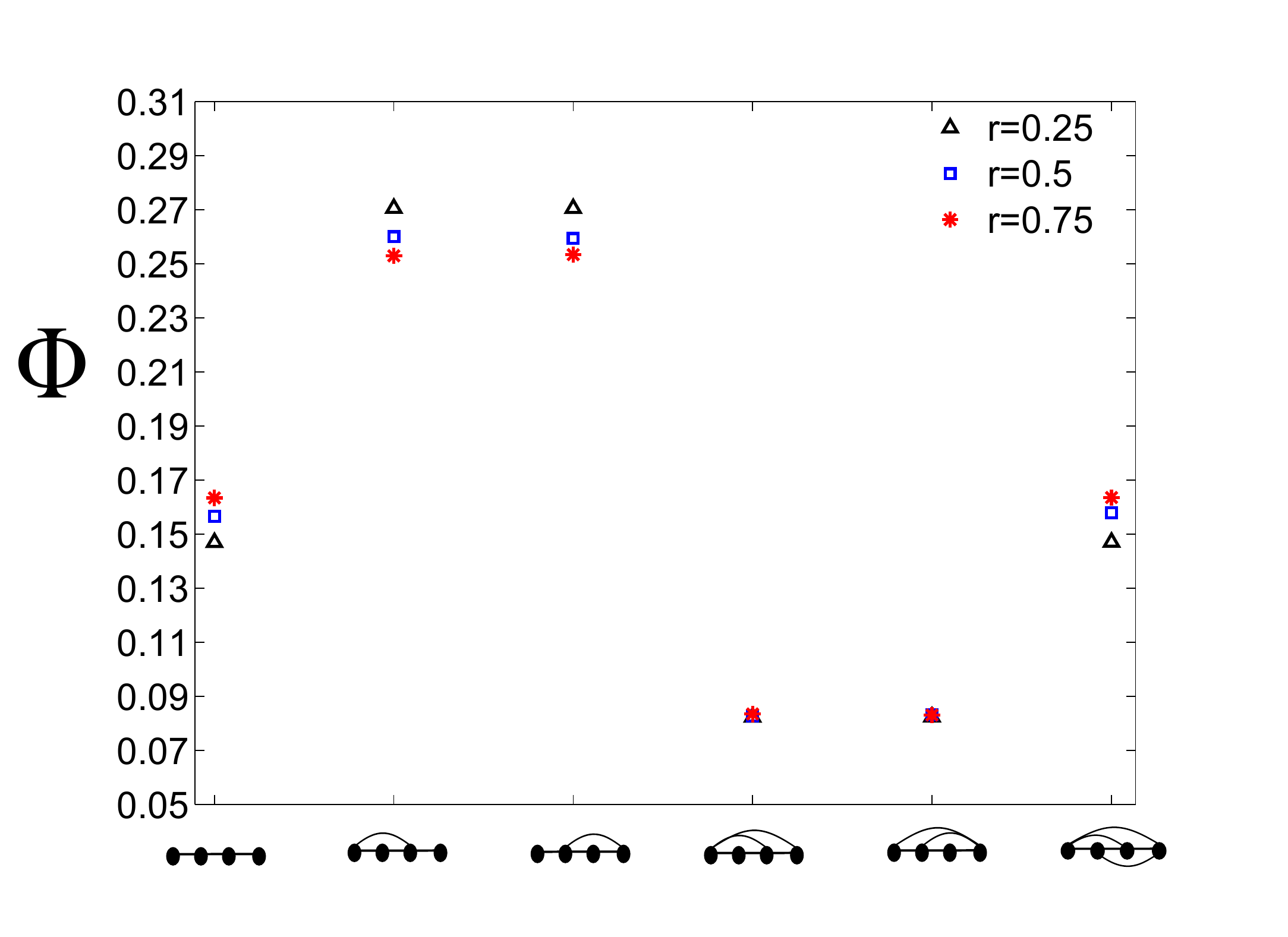}
\caption{Average (50 realizations) frequency of appearance $\Phi$ of VG motifs extracted from the AR(1) processes described by eq.\ref{eq:rednoise}, for different values of the correlation coefficient $r=[\frac{1}{4},\frac{1}{2},\frac{3}{4}]$. Error bars are contained in the symbols; results are in very good agreement with the theoretical expected value.}
\label{fig:1_2}
\end{figure}

\noindent In all these examples, theoretical results are in very good agreement with results obtained with numerical simulations reported in Figure \ref{fig:1_2}.

\subsection{Noise characterization} 
\begin{figure}[h!]
\centering
\includegraphics[width= 8 cm]{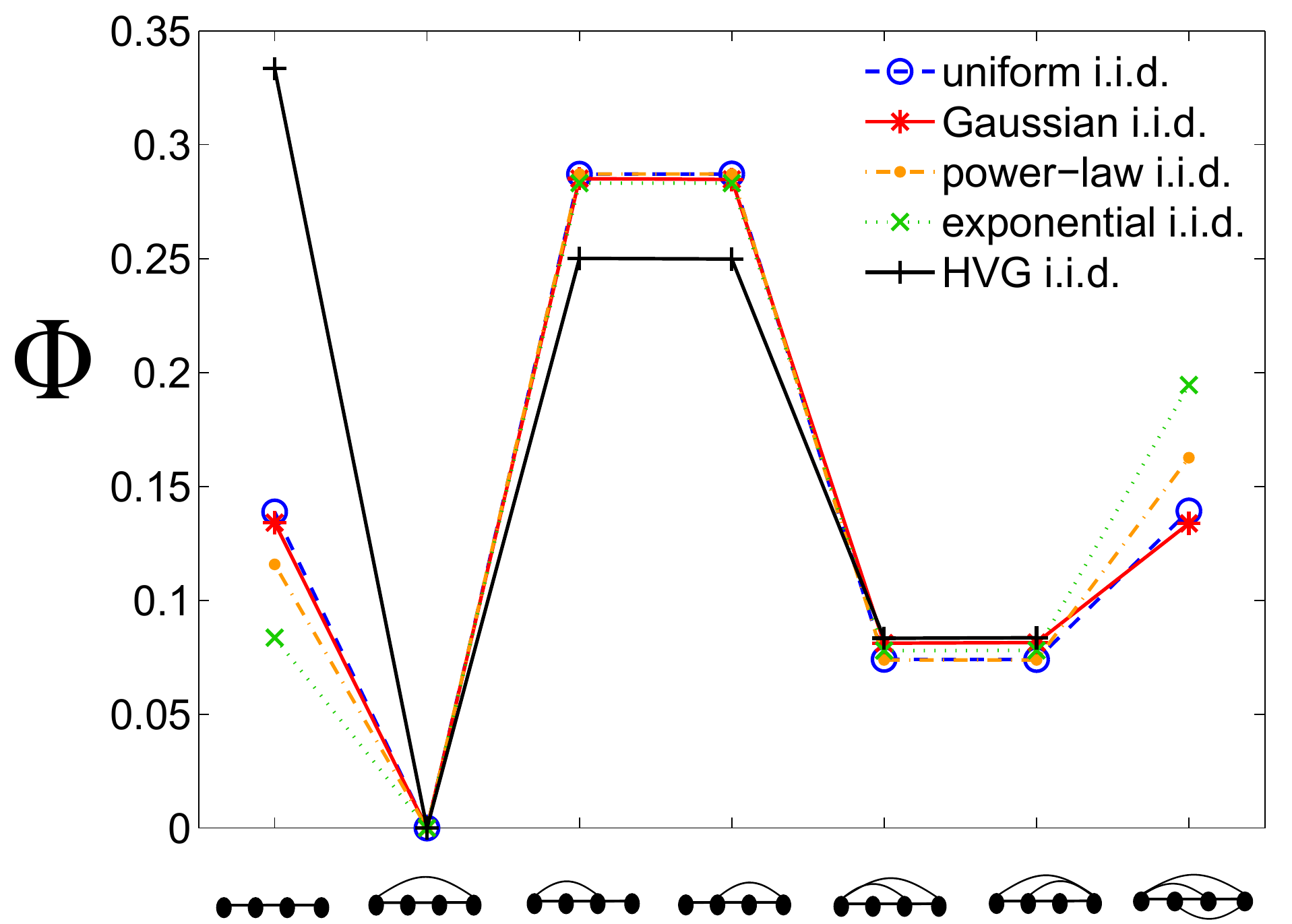}
\caption{Average frequency of appearance $\Phi$ of VG and HVG motifs extracted from i.i.d. series of $N=10^5$ with different marginal distributions; VG motifs are not related with the ranking statistics of the data and able to discriminate the different types of noise.}
\label{fig:2}
\end{figure}

Differently from the HVG motifs, VG motifs statistics does not depend uniquely on the ranking statistics of the data and therefore the VG motif profile could be able in principle to discriminate white noises with different marginals. In the latter sections we have been able to distinguish between Gaussian and uniform white noise. In Figure \ref{fig:2} we summarize the motif frequencies $\Phi^4_m$ of VG motifs forming $Z^4$, extracted from i.i.d. series with different marginals:  
\begin{equation}
\begin{cases}
\text{Uniform}\rightarrow x_i\in[0,1];\quad f(x_i)\sim1\\
\text{Gaussian}\rightarrow x_i\in(-\infty,\infty);\quad f(x_i)\sim\frac{\exp(-x_i^2/2)}{\sqrt{2\pi}}\\
\text{Power-law}\rightarrow x_i\in[1,\infty);\quad f(x_i)\sim x_i^{-k}, \quad k= 2.5\\
\text{Exponential}\rightarrow x_i\in[0,\infty);\quad f(x_i)\sim \exp(-k x_i), \quad k= 2.5\\
\end{cases}
\end{equation} 
In every case we extract series of $10^5$ data. The universal profile obtained for HVG is also plotted for comparison. As expected, motif profiles are different for different marginals. Motifs which are symmetric to each other (3 and 4, 5 and 6) occur with equal probabilities, something that doesn't occur when the series is chaotic (Eq. \ref{log}).

\subsection{Summary}
According to the values obtained for the components of $Z^4$, one can extract some heuristic conclusions:
\begin{itemize}
\item{$\Phi_1$, $\Phi_7$} encode information on the marginal distribution of the process as well as its autocorrelation structure.
\item $\Phi_2$ is null as this motif is not a VG. This is at odds with the HVG case, where this is an admissible motif provided the probability of finding consecutive equal data in the series is finite (e.g. for discrete-valued series).
\item The motifs associated to the pairs ({$\Phi_3$,$\Phi_4$}) ({$\Phi_5$,$\Phi_6$}) have chiral symmetry. In other words,  the motifs associated to $\Phi_3$ and $\Phi_4$ are isomorphic, the correct permutation being $1-2-3-4 \to 4-3-2-1$ (the same holds for $\Phi_5$ and $\Phi_6$). Accordingly, for any process which is statistically time reversible, we expect these probabilities to be equal. Reversible processes include linear stochastic processes (and both white and red noise belong to this family), while non-invertible chaotic processes are usually time irreversible (the fully chaotic logistic map is an example).
Time irreversibility of the process is therefore encoded in these terms.
\item $\Phi_8=0$ as this is not a VG and therefore does not appear (not admissible).
\end{itemize}

\section{Robustness: a comparison between VG and HVG motif profiles}

When dealing with empirical time series, the practitioner usually faces two different but complementary challenges, namely (i) the size of the series and (ii) the possible sources of measurement noise. The first challenge can be a problem when the statistics to be extracted from the series are strongly affected by finite-size effects, whereas for the second one needs to evaluate the robustness of those statistics against noise contamination. For a statistic or feature extracted from a time series to be not just informative but useful one usually requires that statistic or feature to be robust against both problems: it needs to have fast finite-size convergence speed and to be robust against reasonably large amounts of additive noise.\\  
        
In \citep{iacovacci2015visibility} it has been already shown that the HVG motif profile has good convergence properties respect to the series size $N$ and it is also robust respect to noise contamination. Here we explore these very same problems for the case of the VG motif profile and we make a detailed comparison of its performance with the HVG motif profile in a range of situations.

\subsection{Convergence properties for finite size series} 
In general, due to finite size effects, the estimated value of any feature fluctuates and deviates with respect to its asymptotic, expected value. For classical features such as the mean or the variance of a distribution, these deviations are bounded and vanish with series size with a speed quantified by the central limit theorem. The estimation of the motif frequencies can be quantitative effected by finite-size fluctuations and one can even observe missing motifs (motifs with estimated frequency $\Phi=0$) which are not actually forbidden by the process but have not appeared by chance. This situation can be overemphasized in the presence of certain types of measurement noise.\\

\begin{figure}[h!]
\centering
\includegraphics[width= 8 cm]{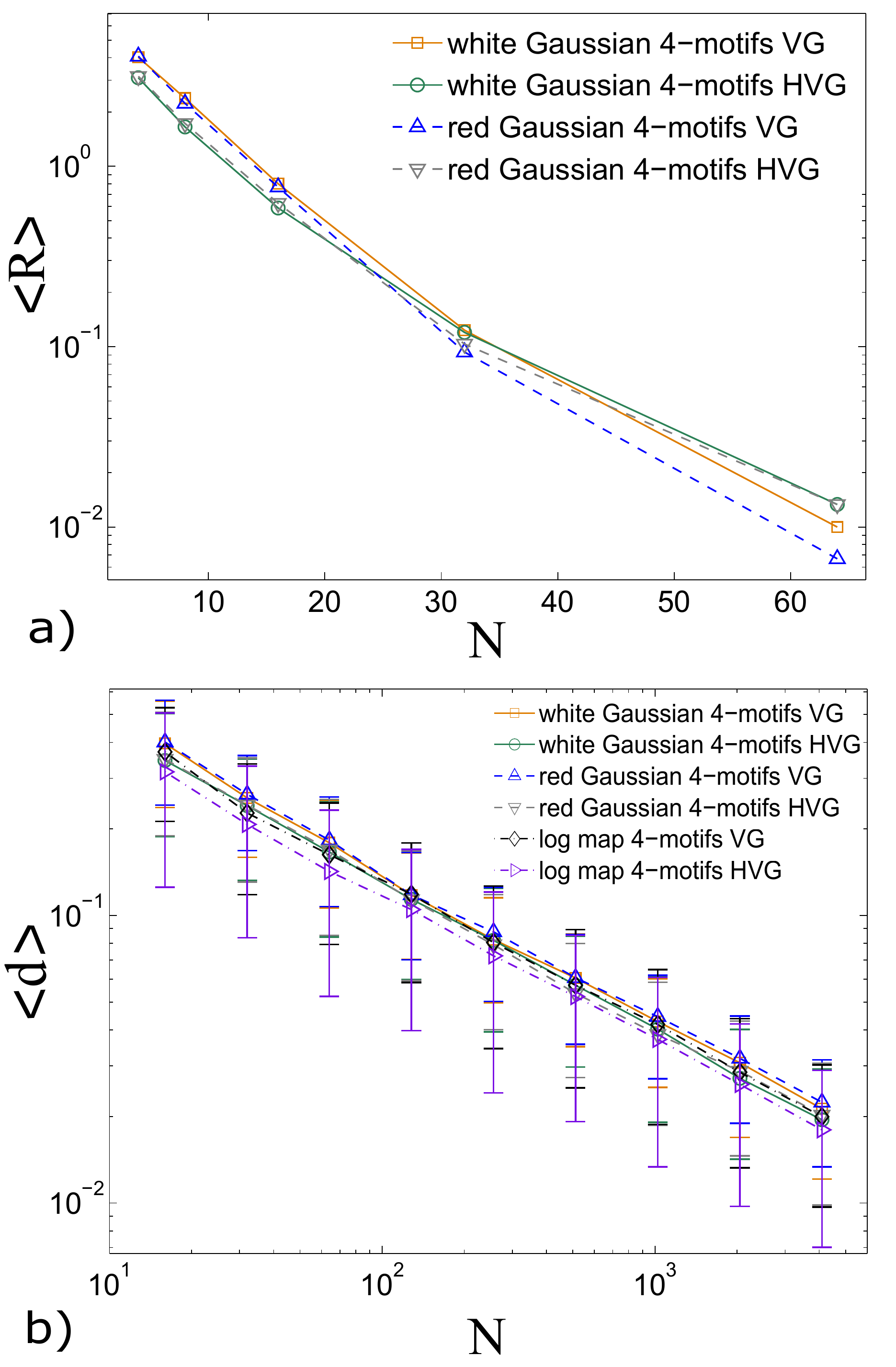}
\caption{Robustness of VG and HVG motif statistic respect to finite series size effects, in the case of Gaussian white noise and colored (red) Gaussian noise. Panel (a): Semi-log plot of the average number of missing motifs $\langle R\rangle$ vs series size $N$ (each point is an average over 300 realizations of the corresponding process). $\langle R\rangle$ decays exponentially to zero, meaning that for $N\sim100$ we can already exclude the possibility of detecting missing motifs due to finite size fluctuations for both type of noise.
Panel (b): Log-log plot of the average distance $\langle d\rangle$ between the observed motif profile and the theoretical profile as a function of the series size $N$ (results are averaged over 300 realizations). $\langle d\rangle$ decreases as a power-law for all the processes considered.}
\label{fig:3}
\end{figure}

\noindent Following an approach analogous to the one followed for the forbidden ordinal patterns in \cite{amigo2007true,carpi2010missing,rosso2012causality}, we first perform a test to study the decay of missing motifs with the series size both in stochastic uncorrelated and correlated processes. In Figure \ref{fig:3} panel a) we plot $\langle R(N)\rangle$, the average number of missing motifs in a series of size $N$ in the case of Gaussian white noise and colored (red) Gaussian noise (for the red noise we consider the AR(1) process with correlation length $r=0.5$ discussed in section III). 
For both types of noise $\langle R(N)\rangle$ decays exponentially to zero and already with a series of about 80-100 data points we can exclude the possibility of detecting missing motifs (for both HVG and VG) due to finite size fluctuations even in the case of correlated noise.\\

\noindent As a second analysis, we explore the convergence speed of the estimated motif profile of uncorrelated and correlated stochastic series and of chaotic series (fully chaotic logistic map) of size $N$ to the asymptotic profile solution given in section III. To do this we define the distance 
between the estimated $4$-motif probabilities $\hat{\Phi}^4_m(N)$ and the asymptotic value $\Phi^4_m=\lim_{N\to\infty} \hat{\Phi}^4_m(N)$. We use $\ell_1$ norm and accordingly define
\begin{equation}
d(N)=\sum_{m} |\hat{\Phi}^4_m(N)-\Phi^4_m|
\end{equation}     
In Figure \ref{fig:3} panel b) we show the trend of $d(N)$ in log-log scale (results are averaged over 300 realizations). The average distance decreases like a power-law for all the processes considered, in agreement with a central-limit-theorem-like argument. For a series of $N=10^3$ points $\langle d(N)\rangle$ is less than $5\cdot10^{-2}$ and the average distance $\langle d_m\rangle$ for each of the single components is less than $10^{-2}$ (not shown).
These results suggest that VG and HVG motif profiles have very good convergence properties and are thus robust against finite size fluctuations. 

\subsection{Robustness against measurement noise} 

\begin{figure}[h!]
\centering
\includegraphics[width= 8 cm]{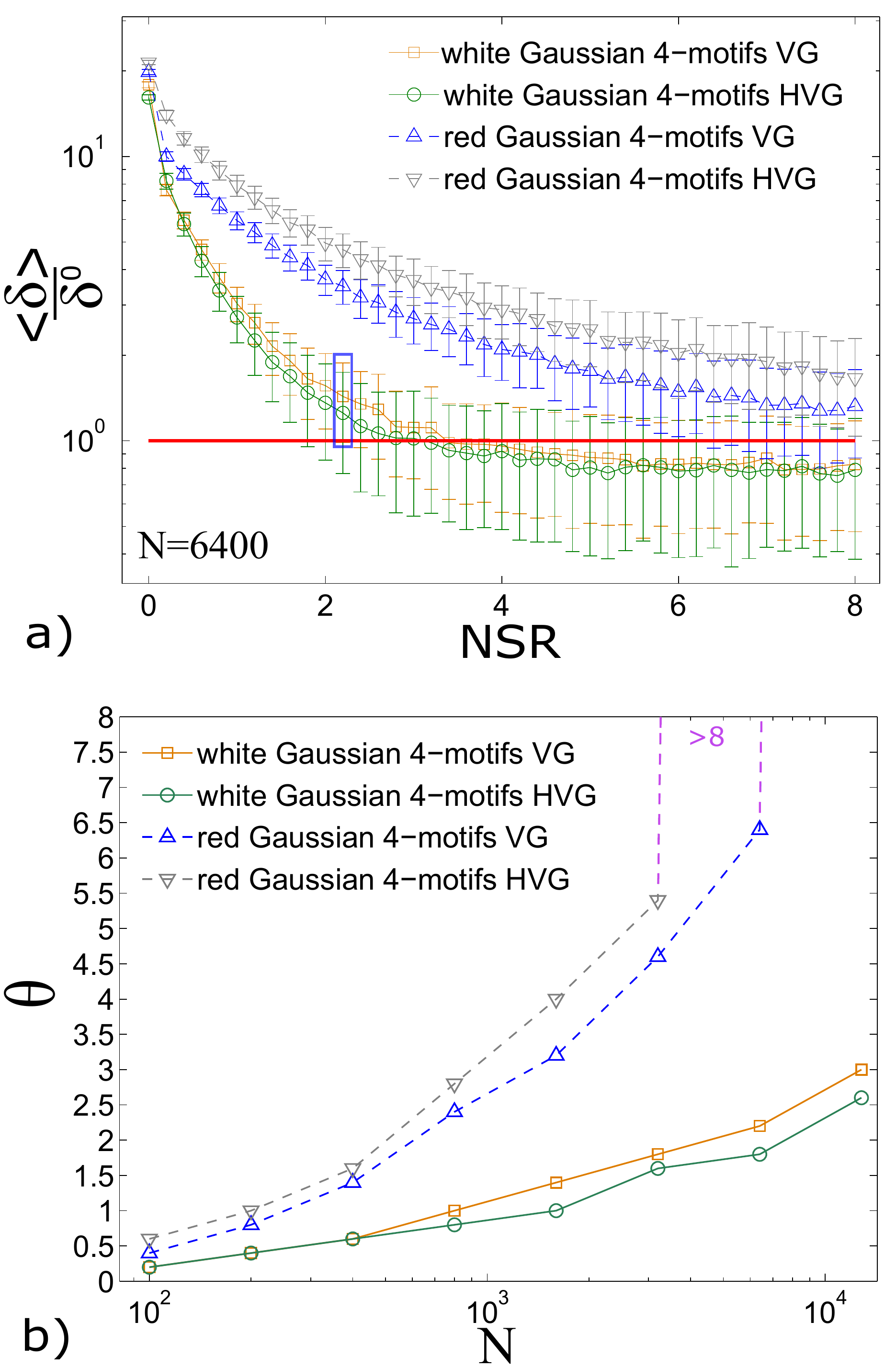}
\caption{Robustness of HVG and VG motif profiles respect to measurement noise. Panel a) the average distance between the motif profile extracted from the polluted chaotic dynamics $X$ (see Eq.\ref{processX}) and the theoretical motif profile of the noise, normalized by the resolution limit threshold $\delta^0$ for different level of contamination NSR (noise-to-signal ratio) at fixed size $N$=6400. When the curves reach the resolution limit (solid red line) we cannot consider any more the motif profile informative about the underlying chaotic dynamics due to the noise effects.
Panel b) the estimated values of the measure $\theta$ in function of the time series size $N$ indicating the maximum amount of NSR for which the motif profile extracted form a single realization of the process $X$ is reliably informative respect to the chaotic signal. The HVG and the VG motif profiles are more robust respect to noise contamination when the noise is correlated (red Gaussian). In this situation the HVG motif profile seems to perform better than the VG motif profile, while in the case of uncorrelated Gaussian noise the VG profile seems in turn slightly more robust than the HVG profile.}
\label{fig:4}
\end{figure}

To test and compare the robustness of VG and HVG motif profiles when the effect of noise contamination combines with the finite size fluctuations we consider the fully chaotic logistic map dynamics $x_t$ polluted with measurement (additive) noise $\eta_t$
\begin{equation}
\begin{cases}
X_t=x_t+\eta_t \\
x_t=4\cdot x_{t-1}\cdot(1-x_{t-1})\\
\eta_t=r \cdot\eta_{t-1}+\sqrt{\alpha}\cdot\xi_t, \quad \xi_t\in {\cal N}(0,1)
\end{cases}
\label{processX}
\end{equation}  
in the two cases where $\eta_t$ is respectively white Gaussian noise ($r=0$) or colored Gaussian noise ($r=0.5$). For both cases $\alpha \in [0,1]$ is the parameter which tunes the noise-to-signal ratio (NSR) of the process defined as 

\begin{equation}
\text{NSR}(\alpha)=\frac{\sigma^2[\sqrt{\alpha}\cdot\xi]}{\sigma^2[x]}
\end{equation}   
where $\sigma^2[\sqrt{\alpha}\cdot\xi]$ and $\sigma^2[x]$ are respectively the theoretical variance of the white Gaussian noise $\sqrt{\alpha}\cdot\xi$ and the theoretical variance of the dynamics (signal) $x$ (note that with this definition we are underestimating the NSR in the case of correlated noise where $\sigma^2[\eta]=\sigma^2[\sqrt{\alpha}\cdot\xi]/(1-r^2)$ ).
The robustness of the observed motif profile $\hat{\Phi}^4_i[X(N,\alpha)]$ for a single realization of the process with given $N$ and $\alpha$ can be defined as the distance between this profile and the theoretical profile $\Phi^4_i[\eta(\alpha)]$ of the noise $\eta$ for the given $\alpha$.  
\begin{equation}
\delta(N,\alpha)=\sum_{m} |\hat{\Phi}^4_m[X(N,\alpha)]-\Phi^4_m[\eta(\alpha)]|.
\end{equation}
With such definition we expect $\delta(N,\alpha)\gg0$ for low values of the NSR (dominant signal, $\hat{\Phi}^4_m[X(N,\alpha)]\simeq\Phi^4_m[x]$) and $\delta(N,\alpha)\simeq0$ for high values of the NSR (dominant noise, $\hat{\Phi}^4_m[X(N,\alpha)]\simeq\Phi^4_m[\eta]$).
Furthermore, $\delta(N,\alpha)$ is affected by finite size effects: if we assume to have few realizations $N_r$  of the process $X$ of small series size $N$, then we expect the variance $\sigma^2(\delta(N,\alpha))$ calculated over the realizations to be high. In particular we have to consider that a resolution limit $\delta^0$ exists, such that when $\langle \delta(N,\alpha)\rangle_{N_r}=\delta^0$ we cannot say any more if the distance we measured is discriminating the signal $x$ from the noise $\eta$ or it is simply due to finite-size effects of the contamination noise $\eta$.
We define this threshold $\delta^0$ as the sum of the standard deviations of the estimated profile components $\hat{\Phi}^4_m[\eta]$ given $N_r$ realizations of the noise process alone    
\begin{equation}
\delta^0(N,\alpha)=\sum_{m} \sqrt{\langle(\hat{\Phi}^4_m[\eta(N,\alpha)]-\langle\hat{\Phi}^4_m[\eta(N,\alpha)]\rangle_{N_r})^2\rangle_{N_r}}
\end{equation}
It is thus convenient to work with the relative distance $\langle \delta\rangle_{N_r}/\delta^0$; when $\langle \delta\rangle_{N_r}/\delta^0\leq1$ we say that the resolution limit for the process $X(N,\alpha)$ -given its $N_r$ realizations- is reached, and $\langle \delta\rangle_{N_r}$ is not any more a reliable indicator.\\
In Figure \ref{fig:4} panel a) we show the quantity $\langle\delta\rangle/\delta^0$ averaged over 300 realizations of the process $X$ for different level of contamination NSR=$[0,0.2,0.4,\dots, 8]$ at fixed size $N$=6400=$100\cdot2^6$ (notice that for $N\geq100$ we missing motifs are not found anymore) respectively for white Gaussian noise and colored Gaussian noise and for the HVG and the VG motif profile.
The red solid line represents the resolution limit threshold for the process. We can see that the HVG and the VG motif profiles are more robust respect to noise contamination when this noise is correlated. In this situation the HVG motif profile seems to perform better than the VG motif profile, while in the case of uncorrelated Gaussian noise the VG profile seems in turn slightly more robust than the HVG profile.\\

\noindent The last step of this robustness analysis is to consider the usual situation where only very few realizations (often a single one) of the same process are available. Our aim is to define a useful indicator $\theta$ which estimates for any given value of the size $N$ the maximum amount of noise contamination level for which a measure $\delta$ computed with only one realization of the process $X$ can be considered somewhat reliable. We define this to be the value of the NSR such that $\langle \delta \rangle-\sigma(\delta)=\delta^0$, 
and thus   
\begin{equation}
\theta(N)=\{\text{NSR}(\alpha): \langle\delta(N,\alpha)\rangle-\sigma(\delta(N,\alpha))=\delta^0\}.
\end{equation}
$\theta(N)$ measures (in units of noise-to-signal ratio) the (statistical) reliability of the motif profile extracted form a single time series of size $N$ of the signal $x$ in the presence of measurement noise $\eta$.\\
In figure \ref{fig:4} (panel a) we plot $\theta(N=6400)$ for white Gaussian noise in the case of VG by considering the $\langle \delta\rangle / \delta^0$ curve marked by orange squares and by taking the smallest value of NSR for which an orange error bar intersect the red line (the blue box highlights the region). Wee find $\theta\simeq2.2$, meaning that when working with a single time series of the process $X$ with size $N=6400$, the $\langle \delta \rangle$ distance measured by using the VG motif profile is reliable up to a level of white Gaussian noise contamination $\alpha$ such that $NSR(\alpha)\simeq2.2$.
In Figure \ref{fig:4} (panel b) we report the estimated value of $\theta$ for the VG and HVG motif profiles in the case of white Gaussian noise and correlated Gaussian noise in function of the series size $N$=$100\cdot2,100\cdot2^2,\dots,100\cdot2^7$ (maximum noise contamination level considered was NSR($\alpha$)=8). We can see that the motif profile is in general a robust measure respect to the combined effect of measurement noise and finite size: working with a single time series of only $3000$ points of the process $X$ we can extract both the VG and the HVG motif profiles and expect those features to be informative respect to the underlying chaotic signal $x$ up to a level of measurement noise for which NSR=1.5 in the case of uncorrelated Gaussian noise and NSR=3 in the case of correlated Gaussian noise.\\       
Also and as observed before (Figure \ref{fig:4}b)), given the case of white Gaussian noise contamination the VG motif profile (orange squares) seems to perform slightly better than the HVG motif profile (green circles). For colored Gaussian noise the situation is the opposite and the HVG motif profile (reversed gray triangles) performs much better (almost a gap of one unit of NSR for $N>1600$) than the VG motif profile (blue triangles). For both type of visibility graphs the motif profile is coherently more robust when polluted with colored noise than with white noise. This is probably due to the fact that white noise breaks up the correlation structure of the signal faster (respect to the size $N$) than correlated noise. It is also interesting that both types of motif profiles are very sensible to the noise correlations although the different nature of the visibility algorithms.

\section{Conclusions}

Sequential visibility graph motifs are small subgraphs where nodes are in consecutive order within the Hamiltonian path that appear with characteristic frequencies for different types of dynamics. This concept was introduced recently \citep{iacovacci2015visibility} and a theory was developed to analytically compute the motif profiles in the case of horizontal visibility graphs (HVGs).
In this work we have extended this theory to the realm of natural visibility graphs (VGs), a family of graphs where the previous amount of known exact results was practically null. We have been able to give a closed form for the 4-node VG motif profile associated to general one dimensional deterministic and stochastic processes with a smooth invariant measure or continuous marginal distribution, for the cases where the variables belong to a bounded or unbounded interval. In the case where the time series is empirical and one does not have access to the underlying dynamics, the methodology still provides a linear time ($O(N)$) algorithm to estimate numerically such profile. We have shown that the theory is accurate and that VG motifs have similar robustness properties as HVG, yet they depend on the marginal distribution of the process and as such yield different profiles for different marginals. This is at odds with the results found for HVGs, where the motif profiles did not depend on the marginals as they behave as an order statistic.\\
The detection of such motifs from a visibility graph extracted from a time series can be seen as a process of dynamic symbolization of the series itself, where the alphabet of symbols is composed by different subgraphs (motifs) which encode information about both data relations and their temporal ordering in their link structure. The deep similarity between HVG motifs and the so called ordinal pattern analysis -which holds mainly due to the fact that HVG is an order statistic- vanishes for VG motifs, which therefore stand as a complementary tool for time series analysis, specially relevant when the marginals play a role in the analysis.

\end{document}